\documentclass[5p]{elsarticle} 

\usepackage{lineno,hyperref}
\modulolinenumbers[5]

\usepackage{amsmath,amssymb,amsfonts}
\usepackage{algorithmic}
\usepackage{graphicx}
\usepackage{textcomp}
\usepackage{xcolor}

\usepackage{ulem}

\usepackage{caption} 

\newcommand{\DLiI}{D^{e^{\text{-}}}}
\newcommand{\cLiINull}{c^{{e^{\text{-}}}}_0}
\newcommand{\cLiI}{c^{e^{\text{-}}}}

\newcommand{\Ds}{D_{\text{s}}}

\newcommand{\cs}{c_{\text{s}}}
\newcommand{\csmax}{c_{\text{s,max}}}

\newcommand{\dSEI}{L_{\text{SEI}}}
\newcommand{\dSEIo}{L_{\text{SEI,0}}}
\newcommand{\VSEI}{V_{\text{SEI}}}
\newcommand{\sSEI}{s_{\text{SEI}}}
\newcommand{\NSEI}{N_{\text{SEI}}}

\newcommand{\USEI}{U_{\text{SEI}}}
\newcommand{\etaSEI}{\eta_{\text{SEI}}}
\newcommand{\kappaSEI}{\kappa_{\text{Li}^+}^{\text{SEI}}}

\newcommand{\jint}{j_{\text{Li}^+}}

\newcommand{\PSOC}{P_0^{\text{SOC}}}
\newcommand{\PSOH}{P_0^{\text{SOH}}}
\newcommand{\QSOC}{Q^{\text{SOC}}}
\newcommand{\QSOH}{Q^{\text{SOH}}}
\newcommand{\RI}{R_{\text{I}}}
\newcommand{\RV}{R_{\text{V}}}

\newcommand\red[1]{{\color{black}#1}}
\newcommand\newred[1]{{\color{black}#1}}

\makeatletter
\def\ps@pprintTitle{}
\makeatother

\begin{document}
	
	\begin{frontmatter}

		\title{Nested State and Degradation Estimation of a Satellite Battery with In-flight Data\\
		}

		\author[affil1,affil2]{Linda J. Bolay}
		\author[affil4]{Omar S. Mendoza-Hernandez}
		\author[affil6]{Eiji Hosono}
		\author[affil7]{Daisuke Asakura}
        \author[affil8]{Sayoko Shironita}
		\author[affil8]{Minoru Umeda}
		\author[affil4,affil5]{Yoshitsugu Sone}
		\author[affil1,affil2,affil3]{Arnulf Latz}
		\author[affil1,affil2,affil3]{Birger Horstmann\corref{correspondingauthor}}
		\cortext[correspondingauthor]{Corresponding author}
		\ead{birger.horstmann@dlr.de}
		
		\address[affil1]{Institute of Engineering Thermodynamics, German Aerospace Center (DLR), Pfaffenwaldring 38-40, 70569 Stuttgart, Germany}
		\address[affil2]{Helmholtz Institute Ulm (HIU), Helmholtzstraße 11, 89081 Ulm, Germany}
		\address[affil3]{Institute of Electrochemistry, University of Ulm, Albert-Einstein-Allee 47, 89081 Ulm, Germany}
		\address[affil4]{Institute of Space and Astronautical Science, Japan Aerospace Exploration Agency (JAXA), 3-1-1 Yoshinodai, Chou-ku, Sagamihara, Kanagawa 252-5210, Japan}
		\address[affil5]{The Graduate University of Advanced Studies (SOKENDAI), 3-1-1 Yoshinodai, Chou-ku, Sagamihara, Kanagawa 252-5210, Japan}
		\address[affil6]{Global Zero Emission Research Center, National Institute of Advanced Industrial Science and Technology (AIST), 1-1-1 Umezono, Tsukuba, Ibaraki 305-8568, Japan}
		\address[affil7]{Research Institute of Energy Conservation, National Institute of Advanced Industrial Science and Technology (AIST), 1-1-1 Umezono, Tsukuba, Ibaraki 305-8568, Japan}
		\address[affil8]{Department of Materials Science and Technology, Nagaoka University of Technology, 1603-1 Kamitomioka, Nagaoka, Niigata 940-2188, Japan}

		\begin{abstract}
            Li-ion batteries are essential for the energy supply of satellites. The accurate estimation of their states is important for the reliable and safe operation in space. This paper introduces a new algorithm for the estimation of SOC and SOH. The multi-timescale algorithm combines Kalman filters and physics-based models for batteries.            
            We use a P2D model combined with a degradation model that describes capacity fading due to SEI growth. The state estimation algorithm combines two extended Kalman filters for the two states evolving on different timescales, with one filter nested within the other one. We test the algorithm with synthetic data as well as with in-flight data from Japanese satellite REIMEI. The algorithm adequately estimates the SOC and SOH in both cases. Furthermore it gives insight into the reliability of the chosen model.
   
		\end{abstract}
		
		
	\end{frontmatter}
	
	

	\section{Introduction}

	Nowadays, lithium-ion batteries are widely regarded as the leading energy storage technology across numerous applications.
	They are indispensable not only in items of everyday use, such as smartphones and electric vehicles, but also in aerospace applications. Especially for satellites Li-ion batteries are ideally suitable, since energy from the sun is steadily available and their reliability in the controlled conditions in space is beneficial for the high safety requirements. 
	
	The Japanese Aerospace Exploration Agency (JAXA) investigate the behavior of Li-ion batteries in satellites. To this end, they launched the small scientific satellite REIMEI in 2005, with a commercial Li-ion battery on board \cite{Uno2011}. 
	The steady environment in space is a good basis for generating long-term cycle data.
	These exceptional data are important to investigate the degradation of Li-ion batteries and to develop and evaluate algorithms that predict the battery behavior.
	
	The requirements for space applications are very rigorous, which is why the devices on board need to be tested extensively and reliable predictions of their long-term behavior is desirable, especially when the mission or the conditions are going to change. In case of the batteries on board, it is important to be able to estimate their state of charge (SOC) and state of health (SOH). These states are not directly measurable. This can be done based on sophisticated battery models combined with mathematical algorithms that account for uncertainties in the data.

	Li-ion batteries are on the market since quite a long time. They have already been extensively studied. In recent years, much progress has been made in developing state estimation methods based on battery models, deploying a broad variety of models and techniques. Depending on the states to be estimated, appropriate models are required to describe the evolution of the states.
	For the state estimation it is common to use filtering techniques, for example based on Kalman filters. 
	
	In his early work on Li-ion batteries, Plett pioneered the application of Kalman filtering techniques for state estimation \cite{Plett2004,Plett2006}. He explored a range of filter types, including the extended Kalman filter (EKF) and the sigma point Kalman filter, combining them with an equivalent circuit model (ECM) to estimate the state of charge (SOC). To address changes in battery behavior over time, he also proposed a dual filtering method capable of tracking both the SOC and time-varying model parameters, effectively capturing cell aging effects \cite{Plett2004b}.

	A variety of models are used to estimate the SOC in batteries. Among them, the ECM is particularly popular in battery management systems (BMS) implemented on microcontrollers due to its simplicity and low computational demands. ECMs are frequently combined with filtering techniques, as demonstrated in several studies \cite{Li2020b,Andre2013,Walker2015,Zou2015,Campestrini2016,Wei2017}. Despite their practical advantages, ECMs are typically limited to the specific operating range for which they were calibrated and offer limited insight into the internal electrochemical processes of the battery.
	
	To gain a deeper understanding of battery behavior, more detailed physics-based models, such as the pseudo-two-dimensional (P2D) model and the single particle model (SPM), are often preferred, even though they are computationally more expensive. One of the early efforts to integrate the P2D model with an extended Kalman filter (EKF) was made by Bizeray et al. \cite{Bizeray2015}, who adapted the model's differential-algebraic equation (DAE) structure following the approach proposed in Ref.~\cite{Becerra2001}. Due to 
	their accurate prediction of cell behavior, such physics-based models have become increasingly popular in Kalman filter applications \cite{Walker2015,Rahimian2012,Stetzel2015,Zou2016,Sturm2018,Bi2020,Li2020a}.

	The estimation of the SOH depends strongly on the underlying battery model and the specific definition of SOH used. Typically, SOH is characterized by the remaining capacity of the cell, but in some scenarios, metrics such as internal resistance or power fade are more relevant. When using an ECM, SOH can be inferred from variations in model parameters, which can be updated over time to reflect cell degradation \cite{Plett2004b}.
	
	For physics-based models, SOH tracking can be handled differently. Some approaches rely on empirical degradation models or adjust parameters to match aged-cell conditions \cite{Rahimian2012,Smiley2018}. Others aim for a more fundamental understanding by incorporating physical degradation mechanisms directly into the model, which enables more robust and predictive SOH estimation \cite{Zou2016,Bi2020,Bolay2022}.
	
	In many practical applications, only the SOC is estimated, while SOH is neglected. This leads to decreasing model accuracy as the battery ages and the model parameters diverge from the true cell behavior. In contrast, some studies focus exclusively on estimating SOH \cite{Bi2020}.
	
	When both SOC and SOH are to be estimated simultaneously, more advanced filtering techniques are required. Even selecting a filter for a single state is nontrivial: while the standard Kalman filter works for linear systems, nonlinear models need more sophisticated methods such as the extended Kalman filter (EKF), the unscented Kalman filter, or particle filters. Comparative studies on these filters can be found in \cite{Plett2006,Walker2015,KleeBarillas2015,Campestrini2016}.
	
	Simultaneous estimation of SOC and SOH is challenging due to their distinct dynamics—SOC changes rapidly, whereas SOH evolves slowly over time. A common strategy is to apply a dual or joint filtering framework based on an ECM \cite{Plett2004b,Plett2006a,Andre2013,Campestrini2016,Wassiliadis2018}. In such methods, both model states and parameters are estimated either through an augmented state vector or via two interdependent filters that update each other iteratively. However, this technique may fail to properly capture the disparity in the timescales of SOC and SOH evolution.

	To evaluate state estimation methods, both synthetic and experimental datasets are commonly employed. Synthetic data offers a key advantage: it provides direct access to the true internal states, which remain hidden in real-world scenarios. This makes it possible to assess whether the estimation algorithm can accurately recover the right values \cite{Rahimian2012,Bizeray2015,Stetzel2015,Zou2016,Sturm2018,Smiley2018}. However, in order to confirm the algorithm’s practical relevance, validation using measured battery data is also essential \cite{Plett2004,Plett2006,Hu2012,Andre2013,Li2020a}.
	
	Experimental data typically stems from controlled laboratory tests, while the use of field data remains relatively uncommon \cite{Li2020b}. The majority of studies focus on electric vehicle (EV) applications, although a smaller number investigate alternative domains, such as aerospace systems \cite{Jun2012,Rahimian2012}.
	
	More advanced methods incorporate multi-timescale modeling. For instance, Hu et al. propose a multi-scale framework that separates the estimation of SOC and capacity and compare it against a dual EKF approach \cite{Hu2012}. Their model, however, is not physics-based and relies on Coulomb counting for SOC tracking. Zou et al. develop a map-based method that accounts for various SOC and input scenarios. This precomputed information can then be utilized during online estimation. Their study, however, is limited to testing their algorithm with synthetic data only \cite{Zou2016}.

	In this work, we develop a new algorithm that estimates SOC and SOH based on a physics-based model and evaluate it with synthetic and field data of satellite REIMEI. The processes in the battery during cycling is described by a P2D model, combined with a degradation model as described in Ref.~\cite{Bolay2022}. For the degradation, we consider the growth of the solid-electrolyte interphase (SEI) as the major reason for capacity fading \red{\cite{Bolay2022,Single2018,Horstmann2019,VonKolzenberg2020,VonKolzenberg2022,Kobbing2023}}. To estimate the states we develop a multi-timescale (MTS) algorithm, where the filtering of the SOC is nested in the SOH filter. This takes into account the nature of the states, which are interdependent while they evolve on different timescales. Our algorithm is first tested with synthetic data, generated according to a satellite application. Then we validate it with in-flight data, spanning around six years of cycling.
	
	In section~\ref{sec:Methods}, we summarize the battery model and explain the multi-timescale algorithm. Next, we describe the synthetic and in-flight data in section~\ref{sec:Data}. And finally, we show the results of the state estimation with both kinds of data in section~\ref{sec:Results}.

	\section{Theory and Methods}\label{sec:Methods}
	
	In this section, we will describe the required methods to estimate the states of a battery. Therefore, we first summarize the model that describes the processes in a battery cell, including the degradation. Then, we develop the algorithm that is able to estimate hidden states of the cell on different timescales. The states that are of interest here are the state of charge (SOC) and the state of health (SOH). Both can be derived from variables of the battery model.
	
	\subsection{Cell and Degradation Model}\label{sec:Models}
	
\begin{figure}[!hb] 
		\centerline{
			\includegraphics[width=0.45\textwidth]{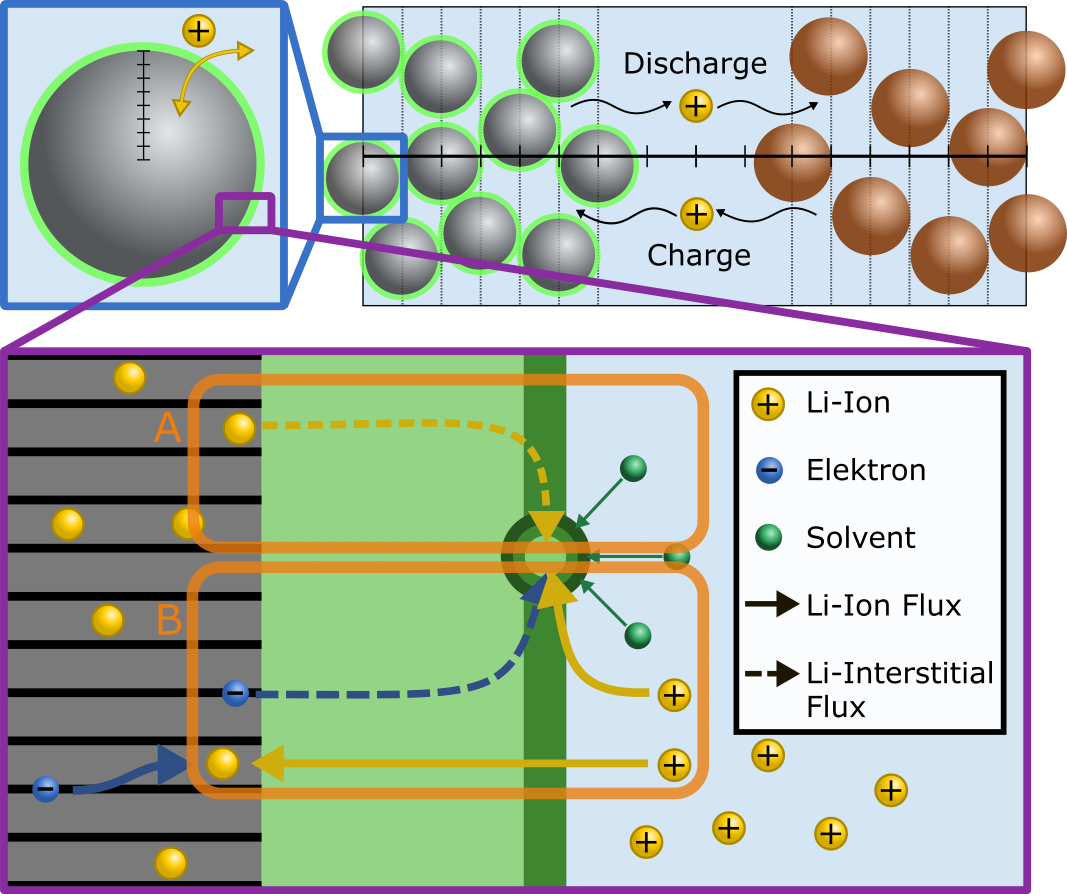}
			\put(-165,200){a)}
			\put(-245,185){b)}
			\put(-245,110){c)}
		}
		\caption{P2D cell model with incorporated degradation model. a) Li-ion transport in electrolyte. b) Reaction at anode particle surface. c) Growth of SEI during storage (A) and during charging (A+B) caused by transport of Li-ions and electrons through SEI. Adapted from  Ref.~\cite{Bolay2022}.}
		\label{fig:models}
	\end{figure}

	The physics-based models used to calculate the states of the satellite batteries have been described in Refs.~\cite{Bolay2022,Schmitt2020}. Here, we summarize the main idea of the models. Fig.~\ref{fig:models} visualizes the different levels of the models and shows the Li-ion transport in the cell, which we describe in more detail below.

	The first model is based on the thermodynamic consistent transport theory of Latz et al. \cite{Latz2011}. In a pseudo two dimensional (P2D) framework the transport of Li-ions in the electrolyte and in the electrode particles, as well as the reaction at the electrode surface, is described. These processes are depicted in Fig.~\ref{fig:models} a) and b). The variable of most interest here, is the Li-ion concentration in the electrode particles $\cs$. It is described by the continuity equation:
	\begin{equation}
		\frac{\partial \cs}{\partial t} = \nabla \cdot (\Ds \nabla \cs) 
	\end{equation}
	with the diffusion constant $\Ds$.
	The P2D model was initially proposed by Doyle et al. \cite{Doyle1993}. With this, the cycling of the cell can be simulated.

	To take the degradation of the cell into account, a second model describes the capacity fading of the cell. The major reason for capacity fading in Li-ion batteries is the continuous growth of the solid-electrolyte interphase (SEI). The SEI is a passivating layer that forms on the anode surface of a Li-ion cell when Li reacts with the solvents of the electrolyte. The mechanism which lets the SEI grow despite its passivating property is assumed to be Li-interstitial diffusion. Electrons diffuse from the anode to the SEI-electrolyte-interface via Li-interstitials in the SEI. There they react with solvents and Li-ions and form new SEI. Single et al. modeled this process during storage of the cell \cite{Single2018}. During charging the growth rate increases as the Li-ion flux in the SEI causes a potential drop which accelerates the electrons. The SEI growth during cycling has been modeled by von Kolzenberg et al. \cite{VonKolzenberg2020} and evaluated with long-term cycling data in Ref.~\cite{Bolay2022}. \red{Horstmann et al. review different models for the formation and growth of the SEI \cite{Horstmann2019}. The interstitial diffusion is compared to solvent diffusion as SEI growth causing mechanism by Köbbing et al. \cite{Kobbing2023}. }
	
	The SEI growth causing mechanism is depicted in Fig.~\ref{fig:models} c). It can be modeled with the following equation for the corresponding flux
	\small 
	\begin{equation}\label{eq:SEI_flux}
		\NSEI  \ = \ \frac{\DLiI \cdot \cLiINull }{\dSEI}  \cdot \exp \left(- \frac{F}{RT} \etaSEI \right) 
		\cdot \left( 1- \omega\cdot\frac{F}{RT} \USEI \right)
	\end{equation}
	\normalsize
	with the SEI thickness $\dSEI$, 
	the interstitial concentration at the anode-SEI interface $\cLiI$, and the reaction overpotential $\etaSEI$. The potential drop at the SEI $\USEI$ is described by
	\begin{equation}
		\USEI =  \frac{\dSEI}{\kappaSEI} \cdot \jint,
	\end{equation}
	with the current density $\jint$ across the interface of electrode and electrolyte (shown in Fig.~\ref{fig:models}~b).
	The adaptable parameters of the degradation model are the diffusion coefficient $\DLiI$, the conductivity $\kappaSEI$ of Li-ions in the SEI, and the migration factor $\omega$. 
	
	The loss of Li-ion inventory reduces the capacity. 
	Another effect of the growing SEI is an increase in the cell resistance. These two phenomenons affect the measurable cell voltage. When the cell gets cycled with a fixed charge and discharge time, the SOC range is set. This range gets shifted to lower values with decreasing capacity \cite{Schindler2017}. Meanwhile, the increase of the resistance leads to higher overpotentials. With this, the thickness of the SEI can be inferred from the cell voltage. 
	
	In this paper, the SOC is derived from the concentration of Li-ions in the anode and the cathode particles and the SOH is described by the thickness of the SEI.

	The state of charge (SOC) at each discretization point within the electrodes is defined as the ratio between the local Li concentration in the solid phase and the corresponding maximum Li concentration of that electrode:
	$ \text{SOC} = \frac{\cs}{\csmax} \cdot 100 \ \%. $
	In this context, we characterize cell degradation by the growth of the SEI layer, and accordingly define the SOH based on this variable.
	The increase in SEI thickness is directly linked to capacity fade, as SEI formation consumes Li-ions that are no longer available for cycling. Consequently, the SEI growth can be quantitatively translated into an irreversible capacity loss $Q_{\text{SEI}}$ via
	\begin{equation}\label{eq:CapacityLoss}
		\partial_t Q_{\text{SEI}} = \frac{AF\sSEI}{\VSEI} \cdot \partial_t \dSEI,
	\end{equation}
	where $A$ denotes the anode surface area, $\sSEI$ is the mean stoichiometric coefficient of Li in the SEI formation reaction, $\VSEI$ is the mean partial molar volume of the SEI, and $F$ is the Faraday constant \cite{Bolay2022}.

	\subsection{State Estimation}\label{sec:SE}

	In many dynamical systems, such as a battery, the states cannot be measured, especially during operation. Therefore, they can only be estimated from the measurable quantities together with a model describing the time evolution of the system. In addition, the data and also the model may be subjected to noise. A commonly used estimation method which also accounts for these uncertainties is the Kalman filter. In the following, we describe the algorithm and some variants. Furthermore, we address cases where the states of a system evolve on different timescales.

	\subsubsection{Extended Kalman Filter}\label{sec:EKF}
	
	The Kalman filter is an algorithm to estimate the hidden state of a linear system. It uses a mathematical model of the system to determine the time update and reduces the error of the state estimate with the measurement in every time step. To estimate the states of nonlinear systems, modifications of the Kalman filter are needed. Here, we make use of the extended Kalman filter. Other filtering techniques, which can be used for nonlinear models, are e.g. the unscented Kalman filter or the particle filter. 
	
	The linear system is described by the time evolution of the system state $x_k$ and its corresponding output $y_k$ at time $t_k$
	\begin{align}
		x_{k+1} &= A_k x_k + B_k u_k + w_k \\ 
		y_k &= C_k x_k + v_k 
	\end{align}
	with the input $u_k$, the system matrices $A_k$, the input matrix $B_k$, the output matrix $C_k$, the process noise $w_k$, and the measurement noise $v_k$ \cite{Kalman1960,Plett2004}.

	In case of a nonlinear model, \red{the discrete time representation of the system is given} as
	\begin{align}\label{eq:KF_nonlinearmodel}
		x_{k+1} &= f(x_k,u_k) + w_k, \\ 
		y_k &= h(x_k) + v_k,
	\end{align}
	where the time evolution of the state and the output of the system are described by nonlinear functions $f$ and $h$. 
	
	\red{The extended Kalman filter (EKF) can be employed for these problems. For that, the nonlinear functions are linearized around the current state and the standard Kalman filter is applied to the linear functions.} For the linearization, the Jacobians $A_k$ and $C_k$ of the functions are determined
	\begin{equation}\label{eq:JacobianEKF}
		A_k = \frac{\partial f(x_k,u_k)}{\partial x_k}\bigg\rvert_{x_k=\hat{x}_{k}^+} \quad 
		C_k = \frac{\partial h(x_k)}{\partial x_k}\bigg\rvert_{x_k=\hat{x}_{k}^-}.
	\end{equation}
	
	The initial state vector $\hat{x}_0$ and state covariance matrix $P_0$ are estimates of the mean and covariance of the system. For the filter, we choose the notation $\hat{x}_{k}^-$ and $P_{k}^-$ for the a~priori estimate of the state and $\hat{x}_{k}^+$ and $P_{k}^+$ for the a~posteriori estimate.
	
	The time update of the state and its covariance $\hat{x}_{k}^-$ and $P_{k}^-$ at time $t_k$ are calculated with
	\begin{align}
		\hat{x}_{k}^- &= f( \hat{x}_{k-1}^+ , u_{k-1}), \\ 
		P_{k}^- &= A_{k-1} P_{k-1}^+ A_{k-1}^T + Q,
	\end{align}
	where $Q$ is the process covariance matrix.
	
	With the updated covariance matrix the Kalman gain matrix $K_k$ can be determined. The measurement update of the state and covariance is then calculated as 
	\begin{align}
		K_{k} &= P_{k}^- C_k^T (C_k P_k^- C_k^T+R)^{-1}, \\ 
		\hat{x}_{k}^+ &=\hat{x}_{k}^- + K_k [y_k- h(\hat{x}_{k}^-)], \label{eq:StateMsrmtUpdt} \\ 
		P_k^+ &= (I-K_k C_k) P_k^-,
	\end{align}
	with the measurement covariance matrix $R$.
	In eq.~\eqref{eq:StateMsrmtUpdt} the term $K_k [y_k- h(\hat{x}_{k}^-)]$ describes the correction of the state. 
	\red{We call this term the "Kalman gain correction" (KGC). We use it as indicator of the sufficiency of the model as we show in sec.~\ref{sec:ResultsSOH} in Fig.~\ref{fig:Results_synthData} and~\ref{fig:Results_inflightData}. The parameters with which the filter can be adjusted to a specific application are the initial covariance matrix of the states $P_0$, the process covariance matrix $Q$, and the measurement covariance matrix $R$.}

	\subsubsection{EKF for differential-algebraic equations}
	
	\red{The nonlinear model can also be described by differential or differential-algebraic equations (DAE). In this case, additional steps are necessary to obtain a linear discrete time representation, which can then be used with the EKF.}
	
	The state space representation of a DAE system is given as
	\begin{align}
		\Dot{x}^d &= g(x^d,x^a,u) \label{eq:StSpMdlDE} \\ 
		0 &= \gamma(x^d,x^a) \label{eq:StSpMdlAE}
	\end{align}
	with the differential equations in eq.~\eqref{eq:StSpMdlDE} and the algebraic equation in eq.~\eqref{eq:StSpMdlAE}. \red{The state vector $x$ is split up into the differential state vector $x^d$ and the algebraic state vector $x^a$. 
		One possibility to obtain a discrete time representation of the system is to use a numerical solver (as for instance \texttt{ode15s} in MATLAB~\cite{MATLAB}). In this case, the function $f$ in eq.~\eqref{eq:KF_nonlinearmodel} represents the solver. The linearization is then carried out as described in eq.~\eqref{eq:JacobianEKF}. 
		
		\begin{figure*}[!h]
			\centering
			\includegraphics[width=0.55\textwidth]{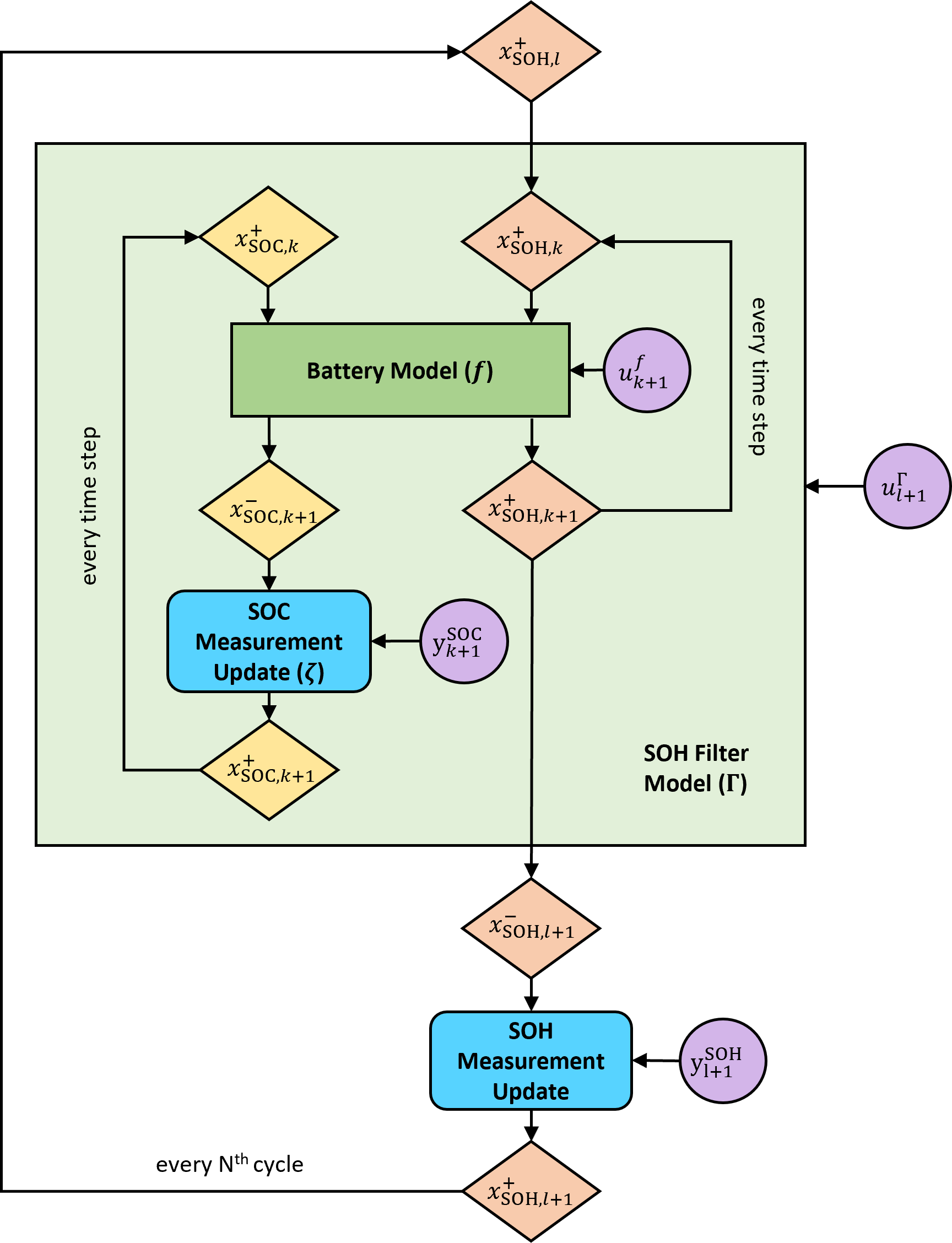} 
			\caption{Flowchart of the multi-timescale filter algorithm. From Ref.~\cite{Bolay2024}.}
			\label{fig:flowchart}
		\end{figure*}
		
		It is also possible to linearize the DAE system first and transform it into a linear state space model of the form
	}
	\begin{align}
		\Dot{x} &= \mathcal{A} x(t) + \mathcal{B} u(t). \label{eq:LinSttSpMdl} 
	\end{align}
	\red{In case of a system of differential equations, the linearized matrices $\mathcal{A}$ and $ \mathcal{B}$ are the Jacobians of function $g$ in eq.~\eqref{eq:KF_nonlinearmodel}. For the linearization of a DAE system the transformation gets more complex. }
	A detailed explanation of the transformation steps can be found in  Refs.~\cite{Becerra2001,Bizeray2015}. 
	As before, the nonlinear functions $g$ and $\gamma$ must be linearized. By rearrangements, the algebraic terms can be expressed by the differential state vector. 
	By adding the stochastic variable $w(t)$, we obtain the stochastic linear state space model
	\begin{align}
		\Dot{x} &= \mathcal{A} x(t) + \mathcal{B} u(t) + w(t). \label{eq:stochLinSttSpMdl} 
	\end{align}

	To discretize this model, eq.~\eqref{eq:stochLinSttSpMdl} gets integrated as described in Refs.~\cite{Becerra2001,Bizeray2015}.
	
	This yields the state matrix $ A_k = e^{\mathcal{A}_kT} $, with $T$ denoting the sampling period. The resulting system can then be processed using the EKF as described in sec.~\ref{sec:EKF}.

	\subsubsection{Nested Multi-Timescale Filter Algorithm}\label{sec:MTS}

	In systems where different processes evolve on distinct time scales, it is important to treat their associated states with separate filters. Without this, there is a risk of over-correcting the slowly changing states. To address this, we develop a nested multi-timescale (MTS) filtering algorithm that integrates two EKFs. In this setup, the EKF handling the fast-changing state is nested within the one for the slow state. A schematic representation of the MTS filter process is shown in the flowchart in Fig.~\ref{fig:flowchart}.
	
	Let the measurement update of the fast state be represented by a function $\zeta$, such that $\zeta(x_k^-) = x_k^+$. By combining the time update and measurement update into a single operation, we define $\zeta \circ f$. We then introduce the operator
	$$
	\Gamma_{\tau(l)} = (\zeta \circ f)^{\tau(l)},
	$$
	which denotes the composition of this operation over $\tau(l)$ iterations, such that $\Gamma_{\tau(l)}(x^+_{l-1}) = x^-_l.$
	Here, $\tau$ is a function that specifies the number of discrete steps required to evolve from time $t_l$, where the slow state is updated by the outer filter, to the next update time $t_{l+1}$.
	
	In order to apply the EKF to the outer (slow) model, a linearization is required, according to eq.~\eqref{eq:JacobianEKF}. The Jacobian of the operator $\Gamma_{\tau(l+1)}$ is computed numerically using a finite difference approximation:
	\begin{equation}
		\frac{\partial \Gamma_{\tau(l+1)}(x_{l})}{\partial x_{i,l}} = \frac{\Gamma_{\tau(l+1)}(x_{i,l}+h)-\Gamma_{\tau(l+1)}(x_{i,l})}{h} 
	\end{equation}
	with a small $h > 0$, where $x_{i,l}$ denotes the $i$-th component of the state vector at time $t_l$. This approximation enables the application of the EKF as outlined in sec.~\ref{sec:EKF}.

	\section{Synthetic and In-flight Data}\label{sec:Data}

	In this publication, we use our MTS algorithm to estimate the SOC and the SOH of a battery system. \red{These inner states are hidden and can not be measured directly. They get estimated} with known or measurable quantities of battery cells. Here, these are the cell current and voltage.
	We test the algorithm with two kinds of data, synthetic and real battery cycling data. 
	\red{Synthetic data are simulated data with added noise emulating a real system with known inner states. 
	}
	The synthetic data are necessary to verify that the algorithm is able to find the true hidden states before it is tested with real data.
	The real battery cycling data originate from the satellite REIMEI of the Japanese Aerospace Exploration Agency (JAXA). In the next section, we describe the in-flight data \red{followed by a description of the synthetic data generation.}
	
	Here, we investigate the degradation of Li-ion batteries under aerospace conditions. So, both the synthetic as well as the in-flight data describe a typical LEO (low Earth orbit) cycling with constant charge and discharge times. With this consistent procedure, the aging of the cells becomes apparent in the end of discharge voltage (EoDV), which decreases due to the increasing cell resistance. The EoDV we assume as additional measurable quantity of the system \red{to use it to estimate the states}.
	

	\subsection{In-flight Data}\label{sec:inflightData}
	
	\begin{figure}[!b]
		\centering
		\includegraphics[width=0.45\textwidth]{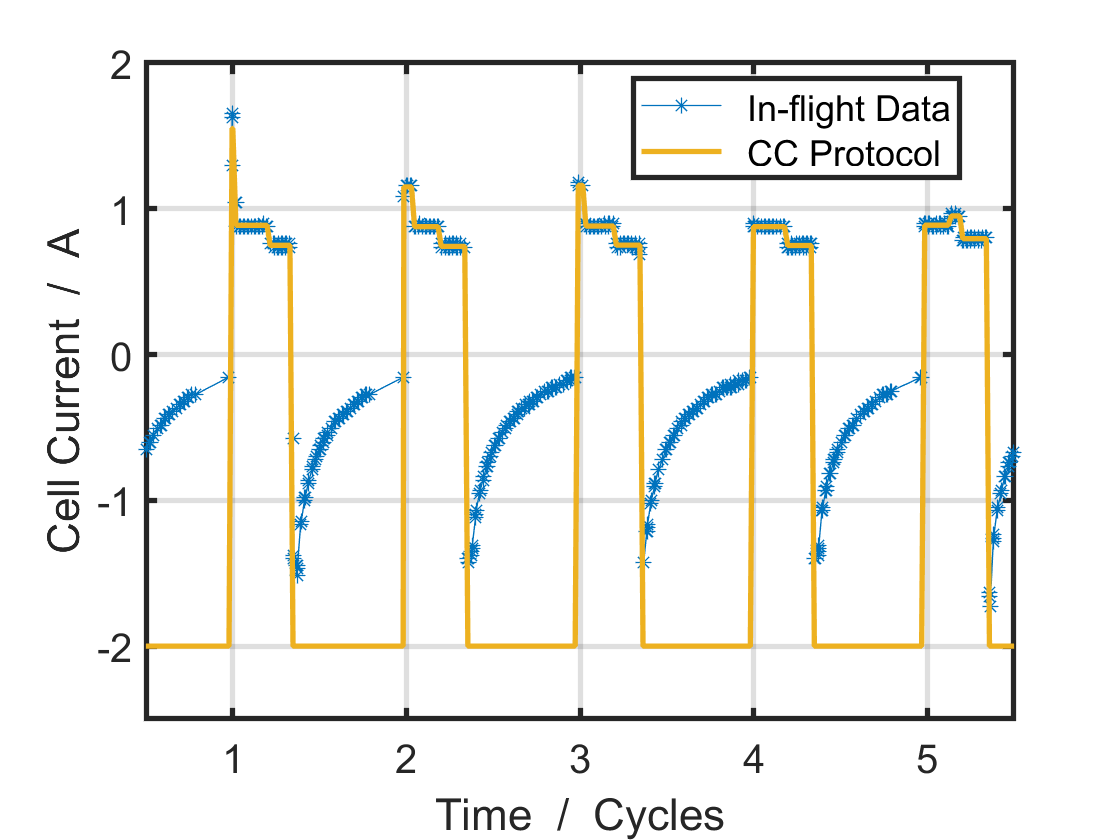} 
		\caption{Five cycles of cell current in-flight data with averaged constant current profiles.}
		\label{fig:InflightData5Cycles}
	\end{figure}
	
	The satellite REIMEI is a small scientific satellite of JAXA. One part of its mission is to investigate the behavior and especially the aging of the on-board commercial Li-ion batteries \cite{Uno2011, Brown2008a}. 
	The satellite was put on a polar orbit in 2005. This is also the start of the cycling of the batteries. \red{The current, voltage, and temperature of the batteries is measured during the whole period of the mission. These data are analyzed to gain further knowledge of the battery conditions \cite{Mendoza-Hernandez2020,Bolay2024}.}

	The battery cycling follows a standard low Earth orbit (LEO) profile. During the sunlit phase of the orbit, the batteries are charged by the solar panels for approximately 63 minutes. In the eclipse phase -- when the satellite passes through Earth’s shadow -- the batteries discharge over roughly 33 minutes to supply power to the loads. Discharge occurs at constant current levels, which vary between 0.4 A and 1.7 A depending on the active loads. Charging follows a constant current / constant voltage (CC/CV) scheme: a current of 2 A is applied until the cell voltage reaches 4.1 V, although in a few cycles this upper cutoff voltage is increased to 4.2 V. The cell temperature is maintained at about 20 °C throughout operation in space. Battery data are recorded at intervals averaging 32 seconds.
	
	In section~\ref{sec:Results}, we will use these constant-value profiles as input for simulating battery cycling and aging. However, since the in-flight data only consist of raw sensor readings without any accompanying metadata, the charge/discharge profiles must first be reconstructed from the noisy measurements. This extensive preprocessing step was applied to cycling data collected between 2005 and 2011.
	
	Fig.~\ref{fig:InflightData5Cycles} shows an example of the in-flight data, where the current 
	measurement points of five cycles are depicted, together with the averaged 
	\red{constant current profiles. We see that there are gaps in the data and that the discharging profiles are not uniform. This emphasizes the necessity of the data processing}. 
	
	\red{Apart from the lack of meta data,} another challenge of the in-flight data is the availability of the measurement points,
	\red{ which} are not acquired continuously. There are recurrent gaps, where only a few data points are missing or gaps that cover several hours. \red{The storage capacity of the satellite covers only 1.5 days for an average measurement interval of 32 seconds. If the data is not downloaded frequently enough, it will be lost. For the gaps, the cycling profiles are determined by averaging the profiles of several time points before and after the gap.
		
		With the cycling profiles generated this way, the in-flight cycling can be simulated and the states can get estimated using the measurement points.}
	
	\begin{figure}[!ht]
		\centering
		\includegraphics[width=0.45\textwidth]{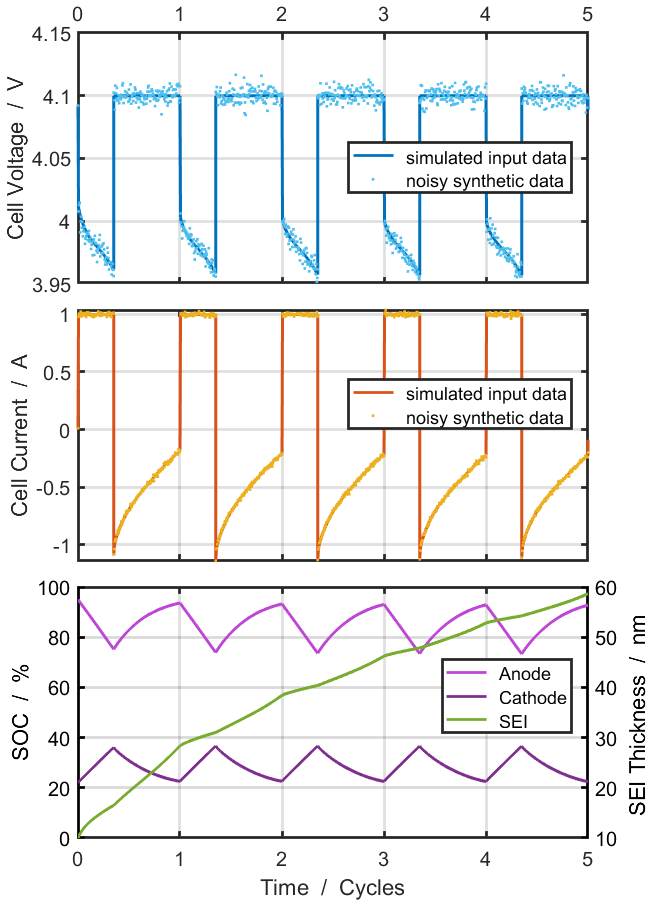} 
		\put(-240,315){a)}
		\put(-240,205){b)}
		\put(-240,105){c)}
		\caption{Synthetic data of five typical LEO cycles. Noise is added to the simulated output data to emulate measured data. a) Cell current. b) Cell voltage. c) Corresponding SOC and SEI thickness. From Ref.~\cite{Bolay2024}.}
		\label{fig:SynthData5Cycles}
	\end{figure}
	
	\subsection{Synthetic Data}\label{sec:synthData}

	To evaluate the performance and accuracy of the MTS algorithm, we employ synthetic data. These data are generated by simulating a representative LEO cycling scenario using the cell and degradation model described in sec.~\ref{sec:Models}. In Ref.~\cite{Bolay2022}, we parameterized this model by using experimental data of the satellite batteries.
	
	To emulate the uncertainty typically present in real measurements, we introduce noise to the simulation output. This allows the synthetic data to closely resemble in-flight satellite data, with the essential difference that both the internal states and the underlying model (including its parameters) are fully known.
	
	The simulated cycling protocol is adapted from the laboratory experiments conducted by Uno et al.~\cite{Uno2011}. At a controlled temperature of 25~°C, the cells undergo discharge at a constant current of 1.0~A (C/3) for 35 minutes, followed by a charge phase using a CC/CV profile: 1.5~A (C/2) until the voltage reaches 4.1~V, sustained for 65 minutes. Detailed model parameters are available in the supporting information.
	
	The simulation spans approximately 4000 cycles, with time steps uniformly spaced every 32 seconds. To emulate the behavior of real data, we add white noise to the simulation outputs -- specifically to either current or voltage signals. The added noise is normally distributed with a mean of zero. The standard deviation is set to 0.005~V for voltage and 0.08~A for current, matching the statistical properties observed in measured in-flight data. Fig.~\ref{fig:SynthData5Cycles} illustrates an example of five such simulated cycles: a) and b) display the noisy voltage and current signals, while c) shows the true inner states.

	\section{Estimating Battery SOC and SOH}\label{sec:Results}

	\red{In this section, we apply the multi-timescale (MTS) algorithm described in sec.~\ref{sec:MTS} to estimate the states of a Li-ion battery.}
	Both, the data and the model are subject to noise and uncertainty. We use the MTS algorithm to cope with the uncertainty in the data as well as in the model, where the uncertainty of the model mainly concerns its parameters. The aim of the filtering process is to get estimates of the state of charge (SOC) and the state of health (SOH) as well as to get insights into the accuracy of the model. 
	
	To proof the validity of the algorithm and to study its performance, in every step, we first use synthetic data and then test the algorithm with the in-flight data.

	\subsection{SOC Filter} 
	
	\red{We start by filtering only the SOC and subsequently we use the MTS algorithm to filter both, the SOC and the SOH. With this, we show that filtering only the SOC is not always sufficient.}
	
	The SOC gets filtered in every time step for which a measurement point is available. 
	\red{In this system, the voltage is considered as output during constant current and the current is the output during constant voltage. For the SOC we choose to perform the filtering during charge, where we use the measurement of the current.} The voltage during discharge could be used equally, but since we also need the voltage for the SOH filtering, we consider this approach more stable. 
	
	For the filter to run properly, the correct filter parameters must be selected. The criteria for the right choice of parameters are that the filter runs stable, that the errors are minimized and that the filtered states are physically sensible. The chosen parameters are the covariance matrices $\PSOC = 10^{-25} \cdot J_n$, $\RI^{SOC} = 1$, and $\QSOC = 10^{-29} \cdot I_n$, where $J_n$ is a $n \times n$ matrix of ones and $I_n$ is the identity matrix of size $n$, where $n$ is the size of the states vector.
	
	\begin{figure}[!hb]
		\centering
		\includegraphics[width=0.45\textwidth]{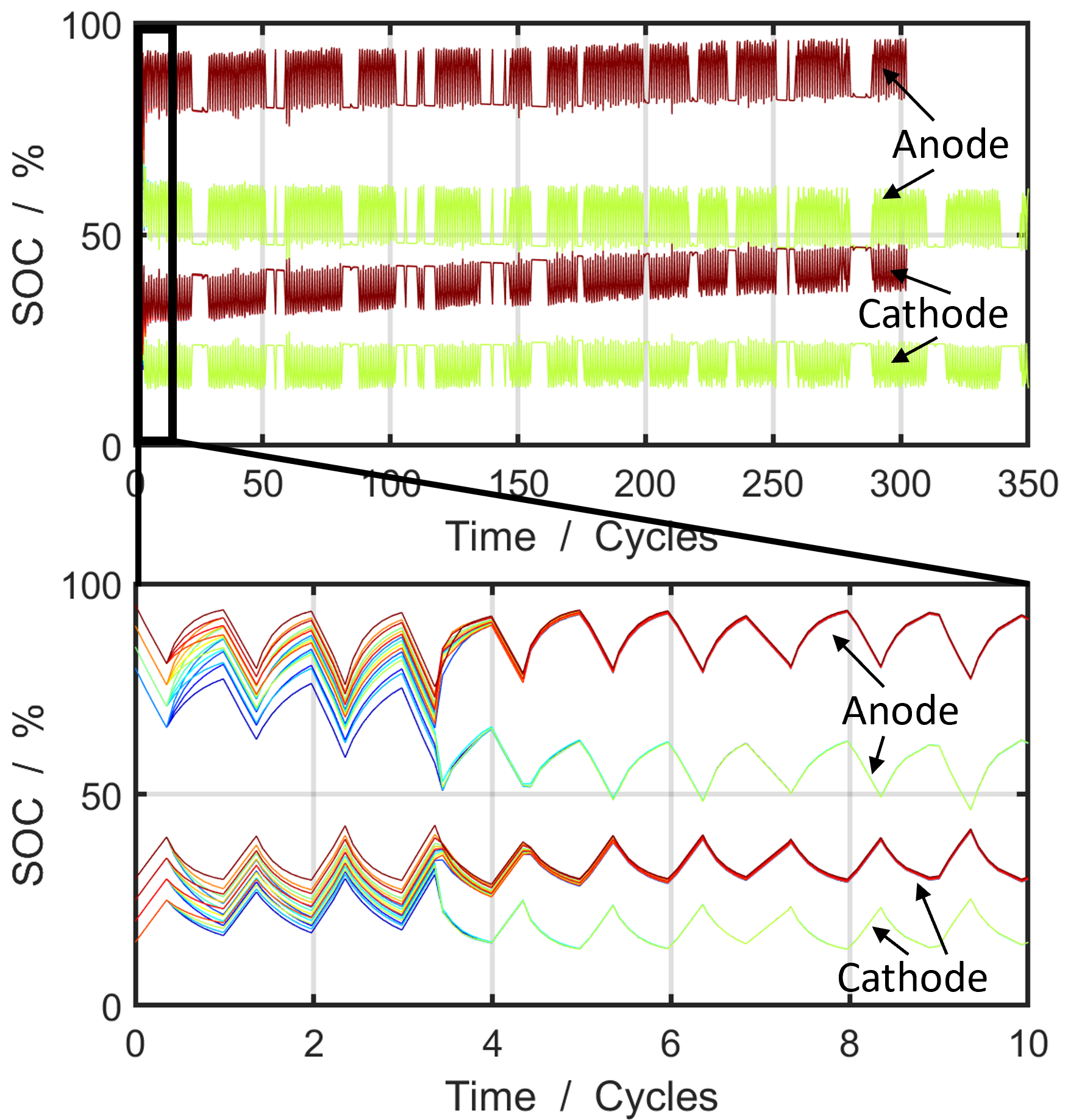}
		\put(-240,240){a)}
		\put(-240,115){b)}
		\caption{SOC filtering with in-flight data. Comparison of anode and cathode SOC trends for different initial conditions. a) Trend of several hundred cycles, where red curves abort after about 300 cycles. b) Zoom-in to first ten cycles. Adapted from Ref.~\cite{Bolay2024}.}
		\label{fig:SOC0_test}
	\end{figure}
	
	With these filter settings the SOC filter is tested with several initial state vectors, where the initial SOC of the anode ranges between 80 and 95 \% and for the cathode between 15 and 30 \%. \newred{The results for the synthetic data can be found in the supporting information.} The choice does not affect the long-term behavior of the filter and all results are very similar. 
	
	\newred{The results for the in-flight data are shown in Fig.~\ref{fig:SOC0_test}, where the SOCs of anode and cathode are depicted for all starting conditions. In this case, some starting conditions lead to an unstable behavior and the simulations abort. In a) the simulation stops after about 300 cycles for some of the simulations, which are plotted in red.} In b) the first ten cycles are zoomed in. It can be seen that in the fourth cycle, when the filter sets in, the SOCs change to smaller values for some of the initial conditions and to larger values for some others. The latter lead to the simulations being aborted. 
	Therefore, we choose the initial SOCs in the range of those that do not lead to instability.

	\begin{figure}[!t]
		\centering
		\includegraphics[width=0.45\textwidth]
		{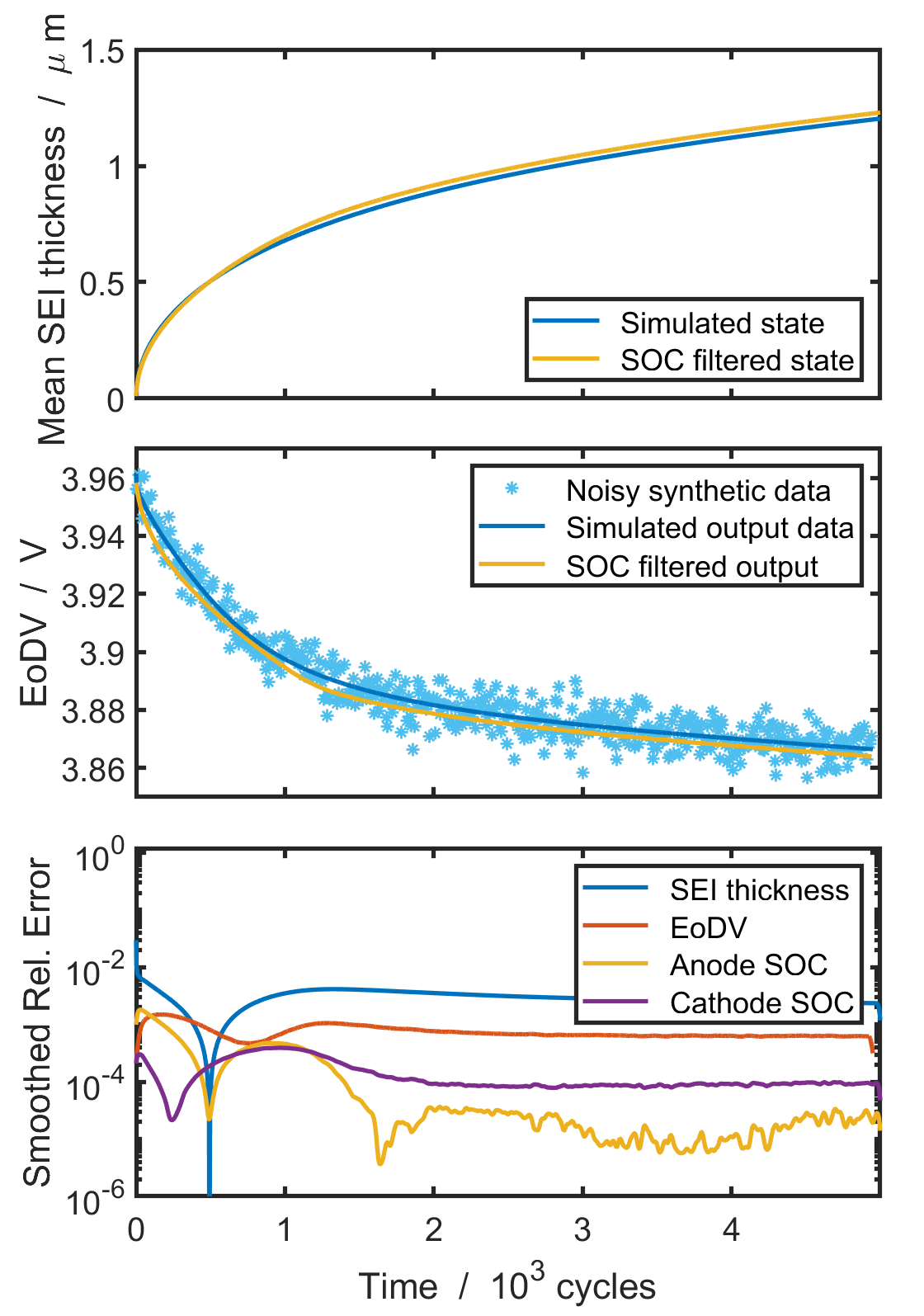} 
		\put(-240,325){a)}
		\put(-240,220){b)}
		\put(-240,120){c)}
		\caption{Results of SOC filtering of synthetic data. a) True and filtered SEI thickness. b) Real, synthetic and filtered EoDV. c) Error of filtered states and EoDV.}
		\label{fig:Results_synthData_onlySOC}
	\end{figure}
	
	\red{To analyse the results of SOC filtering we first consider the synthetic data and subsequently the in-flight data. 
		In Fig.~\ref{fig:Results_synthData_onlySOC}, we show the results for synthetic data. We simulate and filter around 5000 cycles with the typical LEO-charge-discharge-profile described in sec.~\ref{sec:synthData}. The initial SOCs are 80 \% for the anode and 20 \% for the cathode. The figure compares the filtering results of estimated states and calculated outputs with the "true" data known in case of the synthetic data. In a) the SEI thickness, which we use as the variable to describe the SOH, is shown. We see that although the SOH is not filtered here the simulation fits the true data quite well. The simulated EoDV output is compared to the true value and the synthetic measurements in b). The simulated data is in good accordance with the true data. However, there is a small offset right at the beginning, and this will not be corrected during the entire simulation.
		The relative errors of all states and of the EoDV are depicted in c). For all variables the errors are small. Especially for the SOCs the error gets smaller throughout the simulation. These accurate estimates are to be expected since we know the true model in case of the synthetic data. Also, the true initial SEI thickness is known and only the initial SOCs are altered at the beginning to test the filter performance. }

	\begin{figure}[!b]
		\centering
		\includegraphics[width=0.45\textwidth]{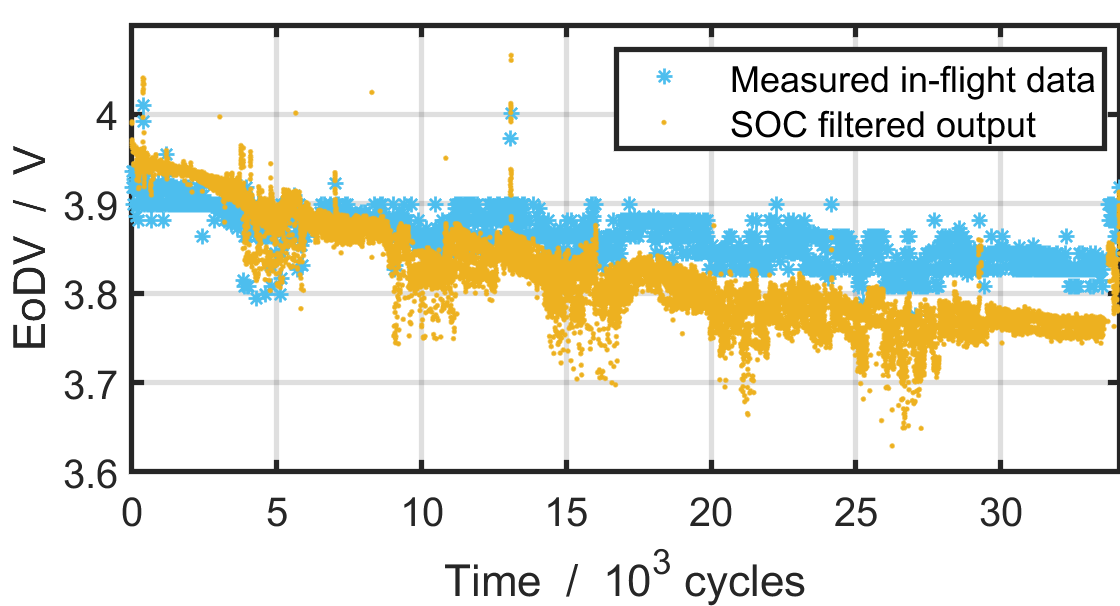} 
		\caption{Results of SOC filtering of in-flight data. Decay of EoDV of the simulation compared to the measured data. Adapted from Ref.~\cite{Bolay2024}.}
		\label{fig:Results_inflightData_onlySOC}
	\end{figure}
	
	\red{When we filter the in-flight data, we cannot achieve the same good results if we only apply SOC filtering. This is depicted in Fig.~\ref{fig:Results_inflightData_onlySOC}, where we show the measured EoDV and the output of the simulation with filtered SOC. We see that the EoDV decreases stronger in the simulation than the measured values, meaning that the cell is degrading faster. 
		In the case of the in-flight data, we do not know the true states, especially not the initial states, and also the true model is unknown. 
		A too strongly deviating initial SEI thickness and too large offsets of the SOC at the beginning of the simulation can lead to deviations in the SEI growth that cannot be corrected in the further course. 
		Also, in our model we only consider SEI growth to describe degradation. But although SEI growth is considered the main mechanism there are many more phenomena taking place in Li-ion cells that contribute to the degradation like cracking of electrode particles or lithium plating \cite{Birkl2017,Hein2016,Mendoza-Hernandez2020,Petzl2014}.
	}
	
	\subsection{SOH Filter}\label{sec:SOHfilter}
	
	\red{For the above reasons it is necessary to also filter the SOH when degradation is taking place. 
	}
	Here, we choose the remaining capacity to represent the SOH, where the capacity is mainly determined by the thickness of the SEI $\dSEI$.
	\red{Li-ions are irreversibly consumed by the SEI and cannot be used for cycling anymore. Eq.~\eqref{eq:CapacityLoss} describes the correlation between the growth of the SEI and loss of capacity. 
	}
	
	The growth of the SEI is changing much slower in time than the SOC. 
	\red{Filtering it in every time step would lead to over-fitting and an unnatural growth behavior.}
	Therefore, the multi-timescale (MTS) algorithm, described in sec.~\ref{sec:MTS}, will be used.

	\red{The filtering of the SOH requires measured values which change accordingly to the change of the SOH. However, in the in-flight data, we do not have measurements of the battery capacity but only of voltage, current and temperature.}
	The progress of the SOH can be observed in the measured voltage, and especially in the end of discharge voltage (EoDV), decreasing over the cells lifetime. 
	\red{The growth of the SEI leads to an increasing inner resistance of the cell. This affects the cell voltage since the overpotential is increasing and the cell voltage is decreasing faster during discharge from the set constant voltage of the charging, resulting in a decreasing EoDV when comparing cycles with the same cycling profile.
		Furthermore, the SOC will shift during aging due to the consumption of cyclable Li-ions, influencing the cell voltage, too.}
	In Fig.~\ref{fig:Results_inflightData_onlySOC} we show the decay of the EoDV for the in-flight data.
	Therefore, we choose the end of discharge (EoD) as the point of time when the SOH gets filtered. The EoDV is then used for the measurement update in the filter. 
	
	Yet, the SOH is not filtered in every cycle since this is too frequent for the slow change of the SOH. We study different frequencies for SOH filtering. A selection of the results is shown in the supporting information. We found that a filter frequency of 20 cycles produces sensible results.
	
	Choosing the time points for the filtering is trivial for the synthetic data. In case of the in-flight data this becomes more complicated as there are gaps in the data. Therefore, the SOH gets filtered every 20 cycles if a measurement point is available. Otherwise, the filtering step gets shifted to the next EoD measurement point. 
	
	The parameters for the SOH filter are chosen analogously to the parameters of the SOC filter. The criteria for selecting the parameters are that they minimize the error of the estimated EoDV, the filter needs to be stable and the estimated $\dSEI$ ought to be physically sensible. These requirements are fulfilled by the covariance matrices $\PSOH = 10^{-20} \cdot J_n$, $\RV^{SOH} = 2.5\cdot 10^{-5}$, and $\QSOH = 10^{-16} \cdot I_n$.

	\begin{figure}[!t]
		\centering
		\includegraphics[width=0.45\textwidth]{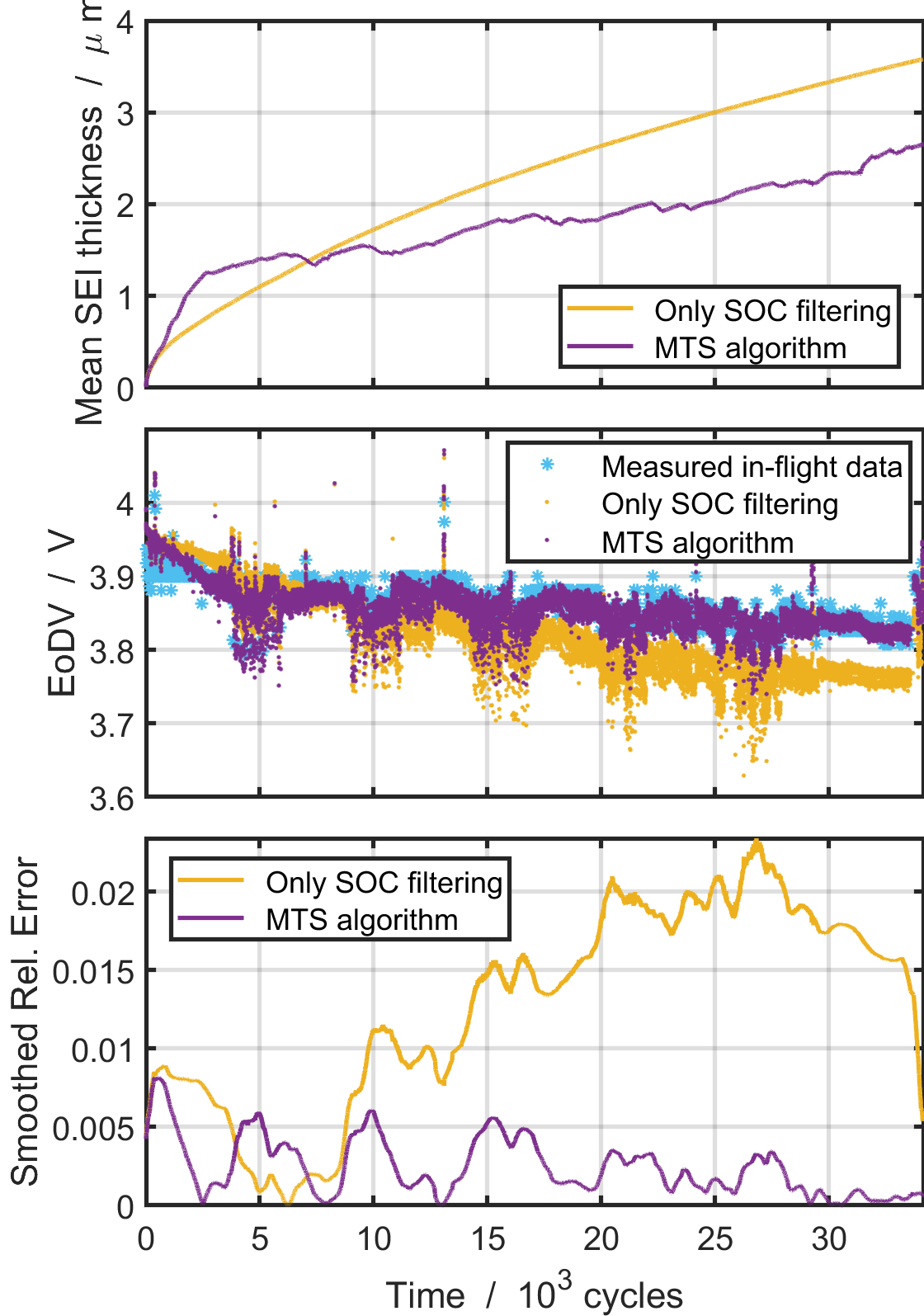} 
		\put(-240,325){a)}
		\put(-240,220){b)}
		\put(-240,120){c)}
		\caption{Results of filtering of in-flight data. Comparison between filtering only SOC and using the MTS algorithm (filtering SOC and SOH). a)~Filtered SEI thickness. b)~Measured and filtered EoDV. c)~Error of filtered EoDV. Adapted from Ref.~\cite{Bolay2024}.}
		\label{fig:Results_inflightData_vglonlySOC}
	\end{figure}
	
	With this setting, the MTS algorithm is applied to the synthetic data and the in-flight data. 
	\red{First, we compare the MTS algorithm to the results of the SOC filtering. Subsequently, we study the performance and features of the algorithm with altered parameters.
		We use the same data as for the tests before, where we filter only the SOC. 
		
		In the case of synthetic data, the results of the MTS algorithm are better than filtering only the SOC. \newred{Details can be found in the supporting information.}
		
		A first result for the in-flight data is depicted in Fig.~\ref{fig:Results_inflightData_vglonlySOC}. The estimation of SOC and SOH using the MTS algorithm is compared to the results when only filtering the SOC. We show the mean thickness of the SEI in a), the EoDV in b), and the relative error of the EoDV in c). The SEI is growing in the same order of magnitude, yet in case of the MTS algorithm the trend tends to be linear after several thousand cycles. We will discuss the reasons for this growth behavior in the next section. For the EoDV we observe a deviation from the measured data at the beginning of the simulation, while they are in good accordance after a few thousand cycles. 
		The error is considerably smaller compared to the SOC filtering after around 7000 cycles.
		This shows that the filtering improves considerably with the MTS algorithm and that the algorithm is able to reproduce the measured data.}

	\section{Discussion of MTS Algorithm}\label{sec:ResultsSOH}

	\red{We investigate the performance of the algorithm when the initial state values are significantly altered
		and how the algorithm behaves when using a different model. 
		Here, we consider variations of our degradation model, where we change the model parameters, leading to an altered SEI growth and thus a different degradation behavior.
		
		To validate the algorithm, we examine the trend of SEI thickness, the error of the EoDV, and if applicable the error of the states.
		Furthermore, we study a feature of the algorithm, the Kalman gain correction (KGC) of the SOH filter from eq.~\eqref{eq:StateMsrmtUpdt}, described in sec.~\ref{sec:EKF}. The KGC indicates how much the filter alters the simulated states to reduce the output error. We will show that this correction can be used to validate the accuracy of the models. We assume that the better the model can reproduce the measured data, the smaller the KGC.
		
		We study several variations, where we scale the degradation parameters and the initial SEI thickness with factors of 0.1 and 10. Here, we examine four variations for which the scaling factors of the parameters and initial states are shown in table~\ref{table:Models}. 
		Model OM is the original model. Model OMTS (original model with thick SEI) also uses the original model parameters, whereas the initial SEI thickness is considerably larger than the one used to generate the synthetic data. Model SDC has a smaller diffusion coefficient and in model SDP all degradation parameters are smaller than in the original model.
		We choose these models since they all produce a small EoDV error but still behave differently. These models illustrate how the KGC can be used as we will discuss later in this section.
		The examination of further models is shown in Ref.~\cite{Bolay2024}.
	}

	\begin{table}[]
		\begin{center}
			\caption[]{Scaling factors for the degradation model parameters and the initial states of four different models.
			}
					\renewcommand{\arraystretch}{1.5} 
					\begin{tabular}{| c || c | c | c | c |} 
						\hline
						Model & OM & OMTS & SDC & SDP \\
						\hline
						\hline
						diffusion coefficient $\DLiI$   & \ 1 \ & 1   & 0.1 & 0.1 \\
						\hline
						conductivity $\kappaSEI$        & 1 & 1   & 1 & 0.1 \\
						\hline
						migration factor $\omega$       & 1 & 1   & 1 & 0.1 \\
						\hline
						SEI thickness $\dSEIo$          & 1 & 100 & 1 & 1 \\
						\hline
					\end{tabular}
			\label{table:Models}
		\end{center}
	\end{table}

	Again, we first examine the filtering of the synthetic data to validate the algorithm in sec.~\ref{sec:MTSSynth}. 
	\red{In this case, we can especially evaluate the capability of the algorithm to estimate the true states when the initial state values are altered.}
	\red{Subsequently, we investigate the MTS algorithm when applied to the in-flight data in sec.~\ref{sec:MTSInflight}. }

	\subsection{MTS Filtering for Synthetic Data}\label{sec:MTSSynth}
	
	\begin{figure*}[!ht]
		\centering
		\includegraphics[width=1\textwidth]{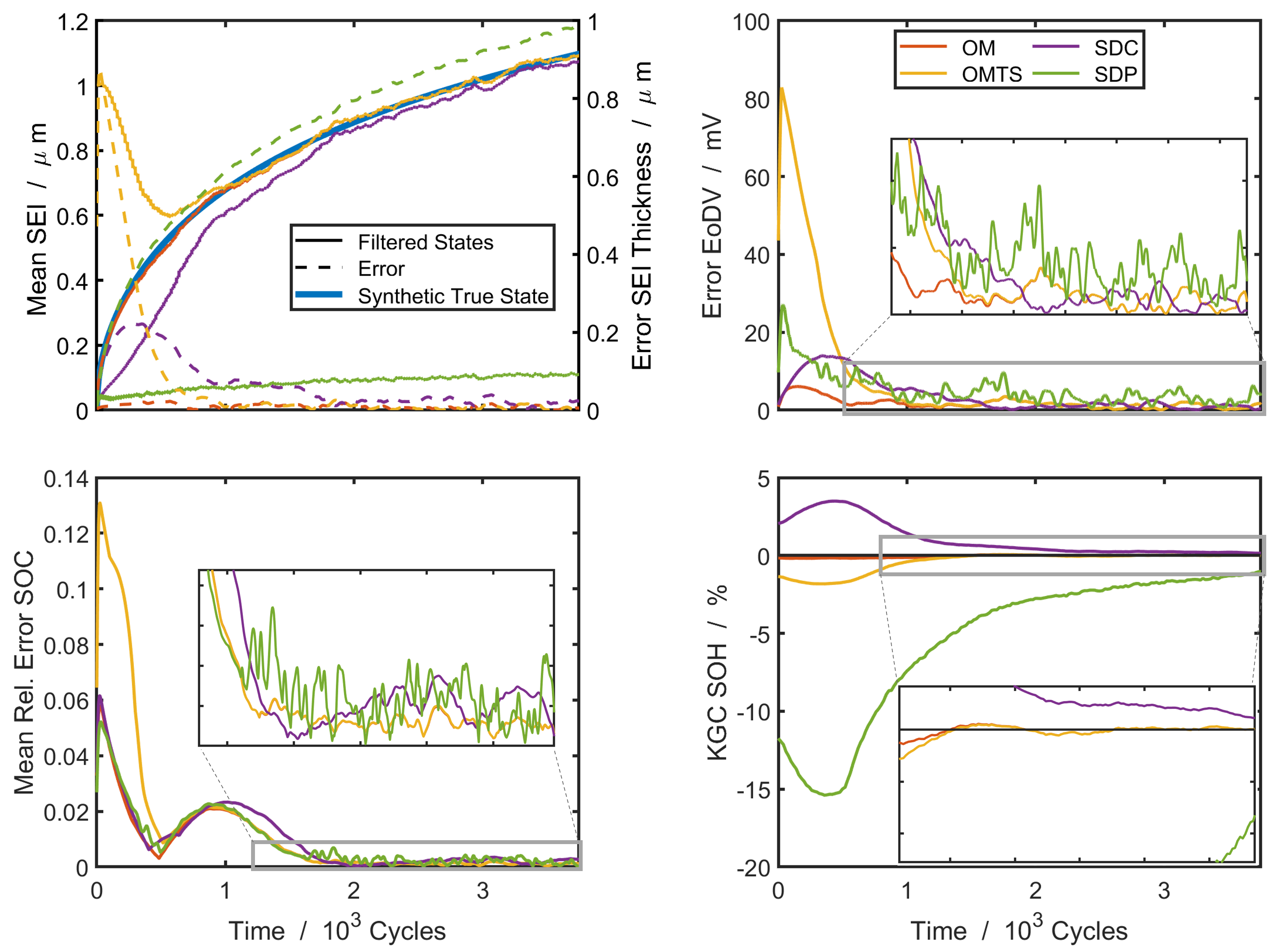} 
		\put(-515,370){a)}
		\put(-240,370){b)}
		\put(-515,190){c)}
		\put(-240,190){d)}
		\caption{Results of the filtering of synthetic data for four different models. See table~\ref{table:Models} for details of the models. Filtered simulation results of the models are compared to the original data. a) SEI thickness and corresponding error. b) Error of end of discharge voltage. c) Mean of relative errors of anode and cathode SOCs. d) Kalman gain correction of the SOH filter. Adapted from Ref.~\cite{Bolay2024}.}
		\label{fig:Results_synthData}
	\end{figure*}
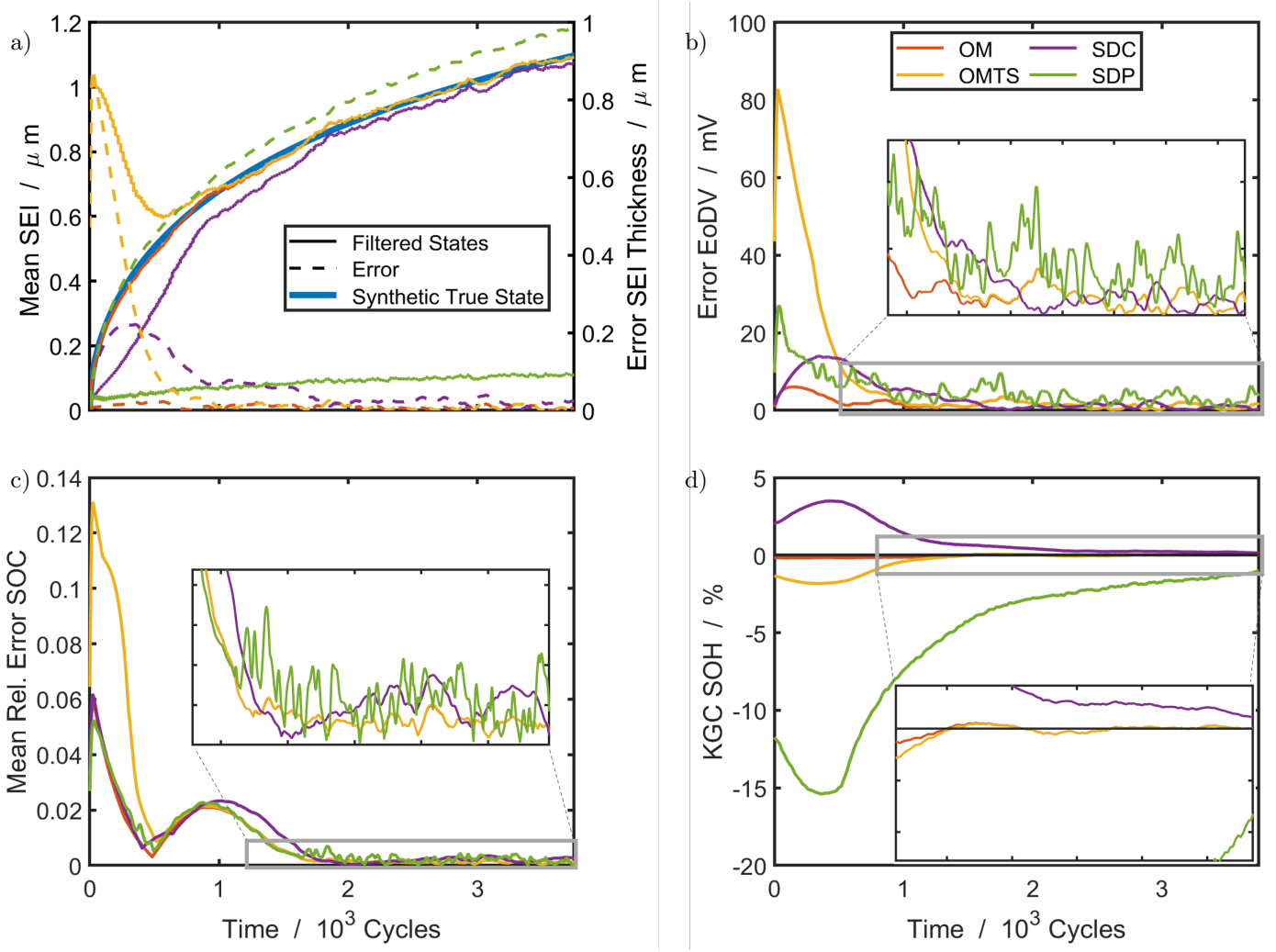

	We show the results of applying the MTS algorithm to the synthetic data in Fig.~\ref{fig:Results_synthData}. 
	\red{
		We simulate around 4000 cycles with the typical LEO-charge-discharge-profile described in sec.~\ref{sec:synthData} and filter it with the MTS algorithm. The initial SOCs are 80 \% for the anode and 20 \% for the cathode. The initial SEI thickness is either 10 or 1000 nm. We show the errors of the states, the EoDV error, and the KGC of the SOH filter, complemented by zoom-ins.
		The error and KGC curves are smoothed in order to better analyze their trends.
		
		In Fig.~\ref{fig:Results_synthData}~a) we plot the trend of the mean SEI thickness compared to the true synthetic states, as well as the error of the SEI thickness. As expected, the original model (OM) has the smallest error. The original model with a thick initial SEI (OMTS) has a large error at the beginning but estimates the true state already after about 500 cycles. For the model with the smaller diffusion coefficient (SDC), the SEI thickness is also in good accordance with the true state but has a slightly larger error than the original models. Only the model with all smaller degradation parameters (SDP) has a large error during the whole simulation and the filter is not able to estimate the true state. 
		In Fig.~\ref{fig:Results_synthData}~b) we plot the error of the EoDV. This output variable is small for all models. After about 500 cycles it only takes on values below 10 mV.
		A similar result is shown in Fig.~\ref{fig:Results_synthData}~c). Here, the relative error of the SOC is shown for all models. We depict the mean of anode and cathode SOC.
		The errors of the SOCs are small for all models. 
		The dip at around 500 cycles is due to the SOC changing from too high to too low values.
		In Fig.~\ref{fig:Results_synthData}~d) we show the Kalman gain correction of the SOH filter for all models. Similar to the trends of the SEI error, the KGC of Model OM and OMTS are very small after a few hundred cycles and oscillate around zero. This means that the filter is correcting the states only slightly. 
		The KGC of the SDC model is also small, although it is positive the entire time, which means that the filter constantly corrects the state upwards to fit the model output, i.e. the EoDV. For the model SDP we see a large negative KGC during the whole cycling period.

		The first result of this study is the capability of the algorithm to estimate the true states. Especially, model OMTS shows that even if the initial state is significantly incorrect, the filter is capable of estimating the true state after several cycles. 
		Furthermore, the algorithm supplies indicators to asses the accuracy of the chosen models. The first indication is given by the error of the EoDV, regulating the filter behavior. If this value is large, the filter is not capable of estimating the states properly with the model in use. But the examples show that this indicator alone is not sufficient and that the model can be wrong even if the filter is able to minimize the EoDV error. The KGC can be used as additional indicator showing how much the filter needs to correct the states to minimize the error. So, a small KGC is an indicator for a suitable model.

		From this, we can conclude that if both, the error of the measurement variable and the Kalman gain correction, are near to zero then both the hidden states as well as the model are correct.
		
		Furthermore, the KGC can be used to improve the model. To do so,  
		the influence of different degradation parameters on the KGC can be studied to eventually choose the parameter set causing the smallest KGC.
	}

	\subsection{MTS Filtering for In-flight Data}\label{sec:MTSInflight}
	
	Finally, we apply the MTS algorithm to the in-flight data. For testing we choose the same models as for the synthetic data. We simulate the first 34,000 cycles of the in-flight data, described in sec.~\ref{sec:inflightData}. With the in-flight data, we do not know the true states nor the true model. The original model used for the simulation of the synthetic data is the one we expect to describe the in-flight data best. In Ref.~\cite{Bolay2022} we explain in detail how the parameters for this model were determined to fit the cell and degradation behavior of the cells used in the satellite. 
	\red{Again, we make use of the EoDV error and the KGC of the SOH filter to analyze the accuracy of the chosen models. }
	
\begin{figure}[!ht]
		\includegraphics[width=0.48\textwidth]{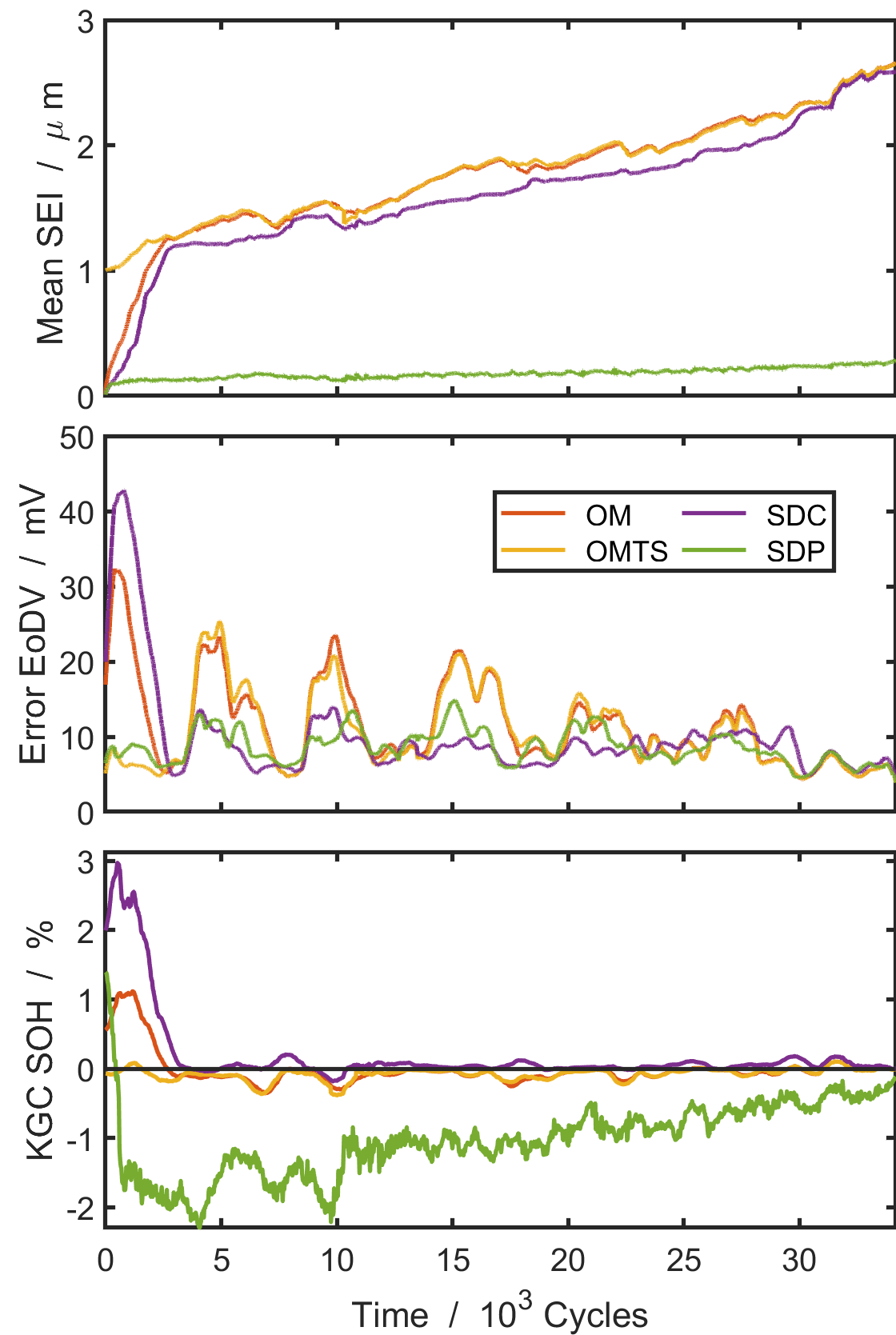} 
		\put(-250,350){a)}
		\put(-250,235){b)}
		\put(-250,125){c)}
		\caption{Results of the filtering of satellite in-flight data. Comparison of four different models. See table~\ref{table:Models} for details of the models. a) Mean SEI thickness. b) Error of the end of discharge voltage. c) Kalman gain correction of the SOH filter. Adapted from Ref.~\cite{Bolay2024}.}
		\label{fig:Results_inflightData}
	\end{figure}

	In Fig.~\ref{fig:Results_inflightData} the results of filtering the in-flight data are shown for the same four models as for the synthetic data. Again, we choose these models for comparison, from all the models we test, because they all have a small EoDV error but show varying behavior in the other quantities. 
	\red{In Fig.~\ref{fig:Results_inflightData}~a) the growth of the mean value of the SEI thickness is shown.}
	As with the synthetic data, the SEI of the SDP model grows significantly slower than that of the other three models, which have a similar growth behavior.
	Especially the thickness of model OM and OMTS is almost identical. Only for the initial approx. 2000 cycles, the thickness deviates a lot as they have different initial values for the SEI thickness. 
	\red{In Fig.~\ref{fig:Results_inflightData}~b) we plot the error of the EoDV. As with the synthetic data, the error is very small for all models.}
	Expectably, the error is almost identical for model OM and OMTS, since these models are identical. 
	\red{The KGC of the SOH filter is shown in Fig.~\ref{fig:Results_inflightData}~c) for all models. We see a similar result than with the synthetic data. The KGC of model SDP is significantly larger than that of the other models. Different to the study with the synthetic data, the KGC of model SDC is just as small as for models OM and OMTS.}
	
	As previously described for the synthetic data, the Kal\-man gain correction can be used to evaluate the accuracy of the models. 
	\red{In case of the in-flight data, we do not know the true model. Considering the small EoDV error, all models would be appropriate. }
	But obviously, model SDP is no appropriate model as the KGC is very large. For the other three models the correction is near to zero, meaning these models describe the cell behavior and the degradation properly. 
	\red{Thus, for an appropriate model, the parameters can be chosen in the range of those with the small KGC. }

	When looking at the trend of the SEI growth of the in-flight data, it is noticeable that already after around 2500 cycles the trend is rather linear. In comparison, for synthetic data we have a square-root-shaped trend. 
	\red{Basically, a linear growth behavior is realistic since this has been observed in experiments for long-term cycling \cite{Keil2019}.}
	There are two major reasons why the trends of our simulations differ, aside from the fact, that the in-flight simulations comprise a significantly larger time span. On the one hand, the cycling protocol for the synthetic simulations differs from the unsteady protocol of the in-flight simulations, where the average discharging current is lower but there are peaks with higher currents\red{, cf. Fig.~\ref{fig:InflightData5Cycles} and Fig.~\ref{fig:SynthData5Cycles}~b). 
		The other reason is that in the case of in-flight data, we try to represent the real circumstances, which include many aging processes, only by SEI growth. And although SEI growth is considered the main contributor to capacity fading, there are processes that take place especially at a later stage of battery life, such as particle cracking, contact loss, and Lithium plating \cite{Birkl2017,Hein2016,Mendoza-Hernandez2020,Petzl2014}. This can lead to a linear aging behavior, which is represented by a linear SEI growth in our model.}

	\section{Conclusion}
	In this work, we describe a new algorithm for estimation of SOC and SOH of Li-ion batteries, which can not be measured directly. The SOC is given by the Li-ion concentration in particles and the SOH is given by the thickness of the solid-electrolyte interphase, the main cause for capacity fading. We use a physics-based model to simulate the cells, combining a P2D model and a degradation model \cite{Bolay2022}.
	
	The algorithm combines extended Kalman filters for state estimation in a multi-timescale framework. This takes into account that the SOH evolves much slower in time than the SOC. To combine the filters, the SOC filtering is nested in the SOH filtering, where the model for SOC simulation together with the SOC filter is considered as the model used by the SOH filter.
	
	The algorithm is tested with synthetic data and with in-flight data of satellite REIMEI, provided by JAXA.  
	
	With the synthetic data, we could show that the algorithm reliably finds the hidden states of the battery. Furthermore, the Kalman gain correction of the filter is an indicator for the reliability of the model, where the correction deviates from zero, when an altered model is used for the simulation and it is consistently close to zero, when the true model is used.
	This is validated with in-flight data of satellite REIMEI, where a model can be found so that the correction is close to zero and the filter also produces reliable estimates of the states. 
	If the correction of the filter is close to zero and the error between simulation and measured data is very small, it can be assumed that the model describes the processes in the battery well. 
	
	A next step could be to add other processes to the degradation model, like plating or the loss of active material. This would be helpful in order to model the degradation of cells especially at later phases of their lifetime more precisely. Also this would improve the safe operation of the cells.

	\section*{Declaration of competing interest} 
	The authors declare to have no competing interests.

    \section*{Acknowledgments}
    \red{This work was supported by the German Aerospace Center (DLR). The authors acknowledge support by the state of Baden-Württemberg through bwHPC and the German Research Foundation (DFG) through grant no INST 40/575-1 FUGG (JUSTUS 2 cluster). This work contributes to the research performed at CELEST (Center for Electrochemical Energy Storage Ulm-Karlsruhe). The authors thank Daniel Möhrle (DLR) for interesting discussions.} 
	
	\bibliographystyle{elsarticle-num}
	\bibliography{StateEstimation}
	
\end{document}